\documentclass[12pt,letterpaper,onecolumn]{article}
\usepackage[latin1]{inputenc}
\usepackage{amsmath}
\usepackage{ mathrsfs }
\usepackage{amsfonts}
\usepackage{amssymb}
\usepackage{braket}
\usepackage{enumerate}
\usepackage{graphicx}
\usepackage{float}
\usepackage [english]{babel}
\renewcommand{\thesubsection}{II\Alph{subsection}}
\usepackage{titlesec}
\titlelabel{\thetitle.\quad}

\usepackage[left=1.00in, right=1.00in, top=1.00in, bottom=1.00in]{geometry}

\author{\\
\\
Horace P. Yuen\\
Department of Electrical Engineering and Computer Science\\
Department of Physics and Astronomy\\
Northwestern University\\
Evanston Il. 60208\\
email: yuen@eecs.northwestern.edu
}

\title{Some  Physics  And  System  Issues  In  The  Security  Analysis  Of  Quantum  Key  Distribution  Protocols}

\begin{document}
\linespread{1}
\maketitle
\linespread{2}

\renewcommand{\abstractname}{ABSTRACT}
\begin{abstract}
\noindent \makeatletter{\renewcommand*{\@makefnmark}{}
\footnotetext{This paper is the preliminary version of one to appear in Quantum Information Processing.}\makeatother}In this paper we review a number of issues on the security of quantum key distribution (QKD) protocols that bear directly on the relevant physics or mathematical representation of the QKD cryptosystem. It is shown that the cryptosystem representation itself may miss out many possible attacks which are not accounted for in the security analysis and proofs. Hence the final security claims drawn from such analysis are not reliable, apart from foundational issues about the security criteria that are discussed elsewhere. The cases of continuous-variable QKD and multi-photon sources are elaborated upon.
\end{abstract}

\section{INTRODUCTION}
In quantum key distribution (QKD), a shared secret bit string $K$ is generated between two users Adam and Babe against possible intercepting attacks from Eve. Security is derived from the quantum effect of information-disturbance tradeoff which is used to bound Eve's possible information on $K$, above which her attack would be detected and below which it can be essentially eliminated. There are a variety of such protocols, many of which have been experimentally studied with strong security claims on the resulting $K$ [1,2]. The term QKD is to be used for this class of quantum protocols only, excluding KCQ (keyed communication in quantum noise) quantum protocols [3] which are not based on information-disturbance tradeoff. A main characteristics of information-disturbance tradeoff protocols is that the transmitted signal be small for ready sensing of the disturbance from an attack. As a consequence it is intrinsically inefficient, lacks robustness, and is infrastructure incompatible. Its important lies in the promise of providing provable near-ideal information theoretic security, as compared to the complexity-based and assumption-dependent security of public key techniques in conventional cryptography.

There are major conceptual and information theoretic foundational problems in connection with the QKD claims, in particular on the empirical operational security guarantees and their relation to the security criterion chosen for security analysis and proofs. Such issues have been discussed in the open literature [3,4,5] and an overview can be found in [6]. They would not be treated in this paper, which is concerned with specific physics and system issues. In particular, we deal here mainly with the questions of whether the mathematical representation of the quantum cryptosystem is adequate for supporting the security claims made on its behalf, irrespective of the security criterion employed or the actual security analysis. The criterion issue and the technical analysis would only come up occasionally as relevant related issues in this paper. Some though not all of the issues in the following have been brought up before with somewhat different explanations.

In section II we discuss the nature of security proof in physics-based cryptosystems, the new issues that come up which are not present in conventional cryptography based on pure mathematical relations and which are often overlooked. Three major problems arise. In addition to one brought to the forefront by detector blinding attacks [7], there are questions of whether all Eve's attack approaches have been covered, especially in the presence of loss, and whether the cryptosystem representation is general enough to capture the intended physical possibilities. In particular, we would explain the error of claiming overall security from the channel mutual information between $A$ and $B$ and between $A$ and $E$, illustrating it in detail in continuous-variable QKD (CV-QKD) in section III. Another illustration is given in section IV on the missing attacks problem, with multi-photon source as concrete example. In section V, we bring out various physics points that we would not be able to go into detail here. The overall moral is that for a complete security proof in QKD, these issues have all to be adequately addressed.

\section{SECURITY  PROOF  AND  QKD SYSTEM  REPRESENTATION}
\subsection{A Proof Or A Conjecture}
A proof of a claimed result is a logically valid deduction of the result from premises that are clearly stated and considered true. Proof is rarely required in science and engineering. It is in fact usually more relevant to demonstrate the result empirically. However, security of cryptographic protocols against all possible attacks cannot be experimentally established, if only because there are an infinite variety of scenarios and also experimental skills and ingenuities. Similar to optimality claims and impossibility statements, they can only be proved or considered as conjectures. Note that one may be able to prove security against a specific class of attacks, but the problem then is to show Eve could not do better from any other attack. 

In the context of cryptographic protocols based essentially on physical effects instead of pure mathematical relations, especially those on quantum effects, this proof requirement engenders a number of new issues not present in usual science and engineering problems, or at least not present with the same degree of seriousness. Are all the physical features relevant to security represented? Is the mathematical representation general enough to warrant the claimed result? Are all 
possible attacks included in the analysis? These ``meta" questions are deep and far from easy to answer, but they affect the status of the security claim profoundly. It is all too easy to claim, explicitly or implicitly by default, that everything is included and one has a proof. We will examine these questions specifically in the following.

\subsection{Completeness of System Representation}
 In conventional cryptography security proofs, all the attacker Eve's possible actions are confined to the mathematical processing of the available information to her that can be completely represented. There is no need to represent her other possible actions. There is, of course, still the issue of side information she may happen to possess, but typically they arise from faulty physical embodiment of the cryptosystem components that can be remedied when found. These problems occur in all uses of mathematical models for real world application, and they just become more precarious in cryptography due to its subtle slippery nature. However, it becomes much more serious in QKD as pointed out long ago [8], in a paper that also first predicted the detector timing attack. This has to do with the microscopic, indeed single photon level of QKD signal detectino. The detector blinding attacks [7] totally breach the then QKD systems with single-photon detection, but not CV-QKD with coherent detection whose photodetectors see a large number of photons. The attacker takes advantage of the more detailed physics and electronics of current single photon detectors which are \textit{not} represented in the cryptosystem model. Specific counter-measures can surely be developed at least in principle to deal with specific such attack, the question is why any detector scheme is safe from similar exploitation. There are two other related issues of cryptosystem representation, generality of the system representation and the incorporation of all possible attacks.
 
\subsection{Generality of System Representation}
In QKD analysis qubits are used extensively, as exactly two-dimensional state spaces of quantum systems. All actual QKD systems that have been considered use photons and relevant qubits are drawn from photon state spaces, which are necessarily infinite-dimensional. The users can restrict their attention in their signaling and processing to whatever subspaces they desire, including just exactly qubits, but Eve should not be so restricted unless one can prove she can't do better. Consider a qubit $H_2$, a subspace of the infinite dimensional $H$ of the optical field under consideration, with the typical four single-photon BB84 states as possible transmitted states. When the photon is coupled into the transmission medium the whole space $H$ is carried with it, one \textit{cannot} chop $H$ off at $H_2$ to deny it to Eve. If Eve wants to do coherent detection measurement on the photon, she can do so readily, a possibility denied of her in a qubit representation of the system. Later in the paper we will give examples why Eve would want to access the whole $H$ in her attack. Here we note that for general security claim a proof is needed to show why $H$ can be replaced by $H_2$ for Eve. That is yet \textit{lacking}.

In fact, this affects the quantitative security level of even ``individual attacks" in which Eve sets up identical probes bit by bit which is entangled to the transmitted qubit. Why is infinite dimensional entanglement not better for her? She can launch an intercept-resend attack by performing a measurement, say coherent detection, that requires the whole $H$ to describe. In probe formulation such an attack would need to be written as an infinite dimensional entanglement. In this connection, it may be noted that the operational security criteria in terms of Eve's various success probabilities should be used finally to validate any security claim against Eve's \textit{optimal} attacks. For example, use of fidelity as approximate cloning criterion is suspect for collective and joint attacks.

\subsection{Including All Attacking Possibilities And Case of Loss}
Foundationally it is a serious problem on how to account for all Eve's information she learnt from the open exchange during the QKD protocol execution [5]. Substantially for the protocol security analysis, it is even more serious how to account for all Eve's possible attacks. This issue has not been addressed in the literature, and claims are often made without giving any reason on why all attacks are accounted for. In this paper we would analyze several such situation, beginning here with the effect of loss on BB84 with single photon sources.

It is generally claimed that the inevitable large optical loss in a realistic QKD situation would affect only the throughput but not the security \textit{level} of single photon BB84, for which not the hint of a proof has been offered. It may be mentioned here, for the purpose of many points in this paper, that the \textit{burden} of proof is on the claimer. Lack of counter-example is definitely not proof, especially in cryptography. That loss may be pernicious in quantum cryptography can be seen from the so-called detection loophole in the testing of Bell inequality on EPR pairs. A classically correlated (or separable) representation can be obtained for a rather small amount of system loss (photodetector quantum efficiency being part of it), say in excess of $18\%$ [9]. This immediately shows the possibility of insecurity in both single-photon and EPR pairs QKD. It is true that no attack from Eve has been specified that utilizes such loophole, but it also shows the usual QKD claim in loss is suspect and a fully spelled out proof with clear scope is needed.

The single photon B92 protocol is well known to be insecure in loss from a ``state determination" attack, and coherent-state BB84 is similarly insecure [10]. This is due to the possibility that the disturbance-information tradeoff principle is affected, at least quantitative level wise, in the presence of loss. By attacking near Adam's transmitter, Eve can do similarly in single photon BB84 [11] and launch a ``probabilistic resend" attack (PRS), with which she deletes the qubits associated with some unfavorable partial or total measurement results she obtained while facilitating transmission of the favorable ones to Babe's receiver. There is no proof why this approach would not change even the individual attack security level, say with approximate probabilistic cloning from Eve.

Note that single photon source is not a sufficient reason for security in loss as B82 shows. While PRS attack cannot lead to security breach in single photon BB84 in the same way as it does in B92, that is \textit{not a proof} of security for BB84 in loss. In fact, there appears to be no reason why the quantitative security \textit{level} of single photon BB84 is not affected \textit{at all} by PRS or other attacks that utilize the possibility of deletion without causing any disturbance. The following shows the quantitative level could indeed be affected.

Let Eve attack a fraction of the single photon BB84 qubits in a round which is below the threshold quantum bit error rate (QBER) in order to avoid detection, by measuring on the ``Briedblatt basis" that determines the bit value with $\sim0.85$ probability of success [1]. That would cause a $0.75$ probability of error in the QBER bit check and that is why only a fraction of the qubits can be so attacked to avoid getting detected. (Note that this is already classified as a joint attack and not even a collective attack, although Eve only needs to set individual qubit probes.) Now Eve may delete the qubits with measured bit value I. Her a priori distribution $p_0(K")$ for the sifted key $K"$ used in security analysis is then no longer the uniform distribution $U$ for $K"$. At typical QBER and loss values, the deviation of such $p_0(K")$ from $U$ is far greater than probable statistical fluctuation. On the other hand, it is not accounted for by just reducing throughput in the security analysis, because such bias in $p_0(K")$ could not have occurred this way in the absence of loss when Eve cannot delete qubits without causing error.

\subsection{Problems of Security Proof from Channel Mutual Information}
Quite a number of security proofs are based on channel mutual information but general security is claimed. This is especially the case in CV-QKD to be analyzed specifically in the next section III. Here we would discuss some problems with the use of channel mutual information for general security proof. To begin with, by ``channel" mutual information $I(A;B)$ we mean as usual that $I(A;B)$ describes the communication capacity of the link between the users A and B. The asymptotic key generation rate is thus often given by the formula, on a far channel use basis
\begin{equation}
R=I(A;B)-I(A;E)
\end{equation}
\noindent where $I(A;E)$ gives the mutual information of Eve's ``channel" on the data transmitted by Adam. It is important to note that this ``channel" mutual information $I(A;E)$ is, generally, not the mutual information $I_E(K)$ that Eve has on some data or key $K$ from her attack. Instead, $I(A;E)$ is obtained by assuming a constant channel may be used to describe how information is derived by Eve, with its usual asymptotic significance given by Shannon's coding theorem as in the case of $I(A;B)$.

The asymptotic significance of such result as (1) immediately shows that it is not accurate for a finite protocol, because neither $I(A;B)$ nor $I(A;E)$ can be achieved as error free rates and the direction of error in the difference between two approximate quantities is not known. Also, (1) does not account for Eve's information from the users' open exchange on error correction and privacy amplification. More significantly, there is no reason for the constant channel assumption to hold under active attack from Eve, especially in QKD with possible entanglement attack. The crucial point is that, as we hope to make clear in this paper, one needs to pay far more attention to the meaning of mathematical formulation and result as they bear on the real world problem of interest. Thus, even more important than the points we just indicated, under an active attack the $I(A;B)$ may lose all its significance because the original channel that yields its value is simply nonexistent.

What is the difference between an \textit{active} and a \textit{passive} attack? It can be characterized as follows, without becoming misleading. A passive attack does not alter the nature of the transmitted signal. It just takes some energy out from it, thus scaling its strength. With the typical large loss in most communication links, one may assume that losing a certain amount of signal energy is already built into the protocol. On the other hand, the transmitted signal is altered in some form in an active attack. Surely if Eve can intercept and launch a passive attack, she can launch an active attack also. A main point of QKD is in fact to turn an otherwise passive attack into an active one from the quantum viewpoint, and hence to detect the presence of the attack if it is judged to threaten security. We next analyze CV-QKD keeping this active-passive distinction in mind.

\section{INSECURITY  OF  CONTINUOUS  VARIABLE  QKD}
In CV-QKD [12,13, 2] the continuous quadrature values of the optical field are used to carry information between the users, which may result in higher key rate and since recently, also thwarts detector blinding attacks from the user's coherent (homodyne or heterodyne) detection. In this section we explain and sharpen the vulnerabilities of CV-QKD under heterodyne-resend attack by Eve which were first indicated in [14]. We restrict to coherent state sources, the case of squeezed states is similar when Eve replaces ordinary heterodyne by ``TCS- heterodyne" [15].    

Typically, a coherent-state with complex-amplitude $x+ip$ (in proper units) is drawn from a symmetric Gaussian distribution of zero mean and variance $V$ for a field mode transmitted from Adam to Babe. The great majority of CV-QKD security analysis start with the channel mutual information $I(A;B)$ and $I(A;E)$ for ``direct reconciliation", and $I(B;E)$ for ``reverse reconciliation" (RR), two different procedures for the users to correct and derive a number of bits from Babe's measurement results. The asymptotic key generation rate is derived from (1) without associated quantitative security levels, which is a typical practice in QKD whose pitfall is explained in [3, App A] and in [5,6], and briefly specified in equ(9) of the next section IV. The fractional signal loss corresponding to the system transmittance $T=|t|^2$ is granted to Eve. In accordance with the above subsection IIE, such an analysis would only apply to passive attacks from Eve and has very little security significance. Surely no general security claim can be based on it. In the following, we'll show how Eve's active heterodyne attack launched near the transmitter of Adam, which can easily be carried out in practice, would cause CV-QKD protocol to be either totally breached or inoperable due to frequent false alarms.

It is important to note that cryptographic security is a decision theoretic problem, not an information theoretic one in the sense of coding and asymptotic performance. Eve has an $N$-ary decision problem on the $N=2^n$ possible key values of an $n$-bit key for her attack [3,6]. In CV-QKD the problem before digitization of $m$ is a so-called ``estimation" problem of identifying a continuous parameter $m$. These problems need to be properly represented for security analysis. We will take $m$ to be the real quadrature parameter that is picked by the user in a particular run. Let $m_B$ and $m_E$ be the homodyne measurement result of Babe and Eve from their respective received parts of the $m$. Then
\begin{equation}
m_B=\sqrt{T}m+n_B
\end{equation}
\begin{equation}
m_E=\sqrt{1-T}m+n_E
\end{equation}
where the additive noises $n_B$ and $n_E$ have
\begin{equation}
var\:\:n_B=var\:\:n_E=1
\end{equation}
If Eve performs a heterodyne measurement near Adam with result $m_E$ and resend it to Babe, we have, in place of (2)-(3),
\begin{equation}
m_B=tm_E+n_B
\end{equation}
\begin{equation}
m_E=m+n_E
\end{equation}
and heterodyne noise $var\:\:n_E=2$. We have absorbed all system loss in the link between Adam and Babe in the ``t" of (5), including receiver loss of any kind in addition to transmission loss. It is illuminating to include Adam's estimate $m_A$ of his nominal choice of the transmitted $m$ including a noise or uncertainty $n_A$,
\begin{equation}
m_A=m+n_A
\end{equation}

In direct reconciliation it is seen from (5)-(6) that Eve has a better (more accurate) version of $m$ than Babe. Notice the channels $I(A;B)$ and $I(A;E)$ of (2)-(3) have been changed to Eve's favor for any $T$ with her heterodyne-resend attack (5)-(6). In RR it is $m_B$ that both Adam and Eve want to get it, and Eve has a better version than Adam even when $n_A=0$. It is easy to quantify this with signal-to-noise ratio or estimation mean-square error, but the representation (5)-(7) is intuitively clear and directly applicable to any reconciliation protocol. Thus, whatever open reconciliation protocol is used between Adam and Babe, Eve would derive at least as much information from it as the users. Indeed, this heterodyne-resend attack possibility has never been addressed in the literature, with only channel mutual information analysis [16-18] that do not account for active attacks.

One common way to increase the users' performance versus Eve's is postdetection selection, which is a classical cryptographic technique [19]. What is significant is that in the presence of loss, postdetection selection can be used by Eve also, in fact near the signal transmitter with a built-in energy advantage to begin with. This belongs exactly to the class of PRS attack we discussed above in connection with QKD security in loss. The precise action from Eve depends on the specific full protocol. In the scheme of [16], for example, Eve can launch a heterodyne attack with resend selection when some thresholds on both quadratures are exceeded. The exact quantitative performance as a function of loss and other parameters need to be worked out to tell the change in quantitative security level, and to see whether the system is thereby totally compromised. It is an analysis that should be included in the security proof, but not. In fact, for the four coherent state scheme in [16] security is totally compromised in large enough loss as is well known [10]. Generally, with postdetection selection by Eve coherent detection near the transmitter may lead to occasional accurate detection of the transmitted state in QKD which will be further mentioned in the next section IV. In such instances, PRS attacks would have to be accounted for and protected against if possible.

It appears security can only be obtained in CV-QKD, if at all, by checking the presence of Eve in her heterodyne-resend attack. However, the checking of Eve's presence in an active attack has not been described in a full CV-QKD protocol. The following considerations show that such checking is not realistically feasible, except perhaps under ideal conditions with exactly known fixed loss and small signal level comparable to single-photon BB84. Let us first fix $t$ to be a precisely known nonrandom number, typically with $T\sim 0.1$ or smaller. The amount of excess noise  from Eve that Adam and Babe would find in their estimation of var $n_B$ is, from (5)-(6), given by $t(n_E +n_A)$ compared to just $tn_A$. To begin with, that is a tiny fraction of $n_B$ they would see for small $t$. Then var $n_A$ is a small fraction of $m^2$ from the way a small $m^2$ or $V$ from 5 to 100 can be physically generated, say by attenuation of a laser pulse and control measurement [14]. This may already mask the $tn_E$ contribution entirely. Equally significantly, the link loss $1-T$ is not known to high accuracy and does not remain constant to within a few percent, either relatively or in absolute value. This is the case not only for the open free space channel but also for the fiber channel with many coupling, alignment, and receiver losses. For such an absolute value of total $\delta T= 0.02$ and $m^2=25$, the uncertain signal level already cannot be distinguished from $n_E$. If threshold for checking is set within such inevitable realistic uncertainty and fluctuation, the protocol would be aborted often even when no attacker is present. Such ``supersensitive" system in the sense of lacking any robustness has never been found to work as real engineering systems, as for example in the case of detection in non-white Gaussian noise and resolution beyond the diffraction limit.

\section{PROPER  REPRESENTATION  AND  MULTI-\\PHOTON  SOURCES}

In this section we will first discuss a very common misuse or misquote of result on the secure key rate in papers of QKD theory and experiment, since the beginning to date. Then it will be connected to the security claim on the use of decoy states for multi-photon sources, in connection with the missing attack problem and the system representation problem.  

\renewcommand{\thesubsection}{IV\Alph{subsection}}

\subsection{Formula For Secure Key Rate}

The following formula for the key rate of the generated key $K$, in bits per channel use for asymptotic $|K|=n$ large, is considered to have been proved since [20] for single photon BB84.
\begin{equation}
R=1-2H(QBER)
\end{equation}
It has been generalized to include many nonideal effects in [21] and is widely quoted as the basis of security claims in various QKD protocols, particularly in decoy states QKD (DS-QKD) [22,23] and measurement-device-independent QKD (MDI-QKD) [24], and in numerous experiments such as [25]. However, (8) was derived under the assumption that CSS codes are used for error correction and privacy amplification, and it is a nonconstructive ``random coding" existence result. It appears no QKD group has actually used CSS codes. For error correction that is because there is no known practical decoding algorithm, and for privacy amplification it is far more complex than linear hashing which is universally employed. Furthermore, there is no quantification of the performance of such CSS codes as finite error correcting code (ECC) and privacy amplification code (PAC) necessary for quantifying their concrete use. There is no claim, and certainly no proof, that (8) applies asymptotically to the LDPC codes widely employed currently in QKD for error correction, or to linear hashing for privacy amplification. Such extrapolation on the applicability of (8) has been treated as proven fact without any mention that it is a mere conjecture.

There are several other major problems with the use of (8) as secure key rate, apart from its asymptotic in $n$ character and the above inapplicability problem. In the first place, it is not meaningful to cite a secure key rate without citing the corresponding security level. In the case of (8), this has been thought permissible because the claim is that under such key rate the security becomes arbitrarily close to perfect as $n$ goes to infinity. That is, $n$ is the security parameter in the claimed ``unconditional security" [26] of QKD protocols. It has been pointed out in [3] and references cited therein that the criterion of Eve's mutual information (accessible information) used in security proofs, $I_E(K)$, does not imply security is arbitrarily close to perfect as $I_E(K)$ goes to zero. Perfect security is given by $K=U$, the uniform random variable, to Eve. It is the \textit{rate} of $I_E(K)\to 0$ as $n$ gets large that determines the asymptotic secure key rate. In particular, it is possible that under  $\frac{I_E}{n}\leq 2^{-l}$ , Eve has a probability of $\overline p_1\sim 2^{-l}$ of getting at the whole $K$. That is, it is possible that
\begin{equation}
I_E=2^{-(\lambda n-\log{n})}\:,\:\:\overline p_1\sim 2^{-\lambda n}\:,\:\:\lambda\ll 1
\end{equation}
In such case, $I_E(K)\to 0$ exponentially in $n$ but $K$ is far inferior to $U$. The situation remains the same when the trace distance criterion is used instead. See [5,6]. There is \textit{no} security parameter in QKD and there is always an exchange between key rate and security level. Such exchange is explicitly dealt with in the approach of [27], for example. Note that the trace distance criterion also removes a major gap in QKD security proof via $I_E(K)$ that led to (8) [2,5].

Next, the side information from ECC and PAC announcement are not accounted for in (8). They have recently been dealt with in [28,29], but explicit evaluation is difficult. Indeed, a formula
\begin{equation}
leak_{EC}= f\cdot n\cdot h(QBER)
\end{equation}
is universally employed to account for the ECC information leak to Eve, with an ad hoc factor $1\leq f\leq 2$ and $h(\cdot)$ the binary entropy function. The PAC leak is accounted for with use of the Leftover Hash Lemma for hashing [30]. These problems are discussed
in [5,6] and references cited therein, and would not be repeated here.

Next we are coming to the representation limitation from which (8) was derived. Qubits are used and thus the full infinite dimensional space available to Eve is not included, as discussed in section IIC. Loss is also not included as discussed in IID. As we noted, it is just commonly claimed (though not in [20]) that loss just affects the throughput but not the security level of single photon BB84, and hence lossy system can be analyzed exactly as a lossless one by merely adjusting the key rate loss. It is also claimed that entanglement protocols with EPR pairs between Adam and Babe is equivalent to single photon BB84, but the precise ``equivalence" is not spelled out and proved, as they are surely not equivalent physically and signaling possibility wise. The equivalence is also less plausible in the presence of loss.

The asymptotic formula (8) is modified to incorporate a number of nonideal effects in [21] heuristically. The results are not derived from a mathematical representation of each new situation. Simple arguments are used which, among other problems, focus on some of Eve's actions and miss out others. Some such problem would be brought up next in connection with multi-photon sources.

\subsection{Multi-Photon Source Problems}
Laser with strongly attenuated output is often used as a concrete source of single photons in QKD, with some corresponding multi-photon probability that leaks the bit value upon photon-number splitting (PNS) attack and basis announcement. This may lead to a total breach of security in loss when Eve can harness near the transmitter a sufficient number of multi-photon signals that would make up the throughput of single photon signals at the receiver. We would not discuss here the physical feasibility of such an attack with lossless channel replacement which is granted in QKD security analysis thus far, and also not the significant possibility that Eve could actually learn a lot about the quantum signal or the bit value from multi-photon counts without lossless channel replacement. From [21] the secure key rate (8) is then reduced by those ``tagged" photons with a simple estimate on its size, which implies a great decrease of the secure key rate which easily becomes zero. 

Decoy states [31,22,23] has been considered a universal method of regaining key efficiency while maintaining the same security as single photon BB84, with typically $S\sim0.5$ close to the single photon level 1. In DS-QKD, multi-photon sources at different mean count $S$ levels of a Poisson distribution with yet unannounced S are transmitted, one level usually taken as the ``signal" and the other ``decoys". The users check the counts for each $S$ afterward. Since Eve does not know which $S$ the signal comes from when she splits the photon, her PNS attack would be caught by the users finding too many counts coming from the higher $S$ signals. A modified key rate formula is then used in lieu of (8) [22,23] to restore much key efficiency. A ``generalized" PNS attack is given in [32] which escapes detection by any DS scheme, and which introduces a large number of split photons with typical parameters that can, however, be dealt with by proper privacy amplification as all tagged photons are so claimed in [21,22].  

As pointed out in [32], there is a serious missing attack problem in such claim. Eve knows she would find out exactly the quantum states of the signals in the bit positions she has succeeded in splitting out a photon for herself. She may want to entangle the remaining photon from a multi-photon signal, by itself or in conjunction with other ones, to the single photon signals she could not split out. After the basis announcement she may gain information from measuring on her probe that has been so entangled, via the additional knowledge of the split out photons. By totally separating the pile of multiple photon signals and the single photon ones, such attacks have not been accounted for in the literature. In fact there are a variety of attacks that are open to Eve due to the presence of such sure to be known bit positions, including ones from the known structure of the ECC (without knowing the specific ECC). A proof or rigorous estimate is needed to account for such attacks.

There is a major error of system representation in taking the density operator of each signal from an attenuated laser to be diagonal in the number state basis. Assuming pulse to pulse phase randomization has been carried out on the source, within a laser pulse the phase is definite though unknown. That is why it is a laser. The signal pulse is in a coherent stare of an unknown phase $\phi$, and one cannot randomize the unknown phase, surely not by a uniform phase, before one decides on what is to be done on the signal. There is a difference between the following two states,
\begin{equation}
\rho_1=|\sqrt{S}e^{i\phi} ><\sqrt{S}e^{i\phi}|,\:\:\:\:\:\:\rho_2=e^{-S}\sum_n\frac{S^n}{n!}|n><n|
\end{equation}
In particular, one can coherently split the unknown phase pulse $\rho_1$ by a beam splitter with any proportion of energy between the two output ports. In the representation $\rho_2$ such a splitting can only be done in units of photon number, very large relatively in this case, and only ``randomly" in contrast to coherent beam splitting.

Coherent beam splitting opens up a large number of possible attacks to Eve. Essentially she can split off up to nearly the entire signal to gain information before she decides on what to do with the remaining pulse, which she can present to the receiver or delete. For example, she may try to discriminate between the different $S$ levels from her split off copy. If she succeeds with a good probability, she would defeat decoy states from such knowledge. With the often suggested $S\sim 1$ for the ``signal" and $S\sim 0$ for the decoys, such probability is sizable and a detailed statistical analysis must be carried out to assess the quantitative significance of such attacks.

Indeed, coherent beam splitting offers many other attack possibilities as discussed in section II. For example, a PRS attack by splitting the beam and heterodyning separately on the two bases used, polarization or otherwise, may lead to near determination of the state after postdetection selection. It is very complicated to formulate Eve's optimal attack even if no probe and just measurement-resend is involved. It is an optimization problem with crucial information to be obtained later from the users' open exchange, and also with the possibility of discarding the unfavorable results up to a certain threshold. In fact just the following optimal detection problem has not been addressed or formulated: Given a threshold fraction of measurement results on the incoming signals can be deleted, what is the optimal average probability Eve can obtain on the bit, or state, from whatever measurement and decision rule? Such problem has not been dealt with generally in either classical or quantum theories. It appears they need to be before some complete QKD security proofs can be carried out.

\section{SOME  OTHER  ISSUES}
In this section we will bring up some other related physics and system issues in QKD security proofs. First, in the QBER estimate the users need to transfer the counting on the checked qubits to the sifted key $K"$. This step is rarely addressed, and not correctly till [33]. Counting of classical quantities is used in [34] for such purpose, which underestimates the fluctuation that results from quantum qubits, the different states of which do not correspond to different colored classical balls in ``classical random sampling" [34]. A qubit state, perhaps disturbed by Eve or by system imperfection, may show a result corresponding to another state except ones orthogonal to it. Similar classical counting is used in the security proof of [27] (in Suppl. 2 of the arXiv version). Quantitatively, [33] shows that the probability exponent is reduced by a factor of 2 when correct quantum counting is used in lieu of classical counting.

Secondly, a classical permutation has been used to impose symmetry on the qubits for joint attack security proofs. Such symmetry cannot be so obtained. The specific permutation employed is openly announced, if only because too much secret bits would be needed to hide it from Eve. Thus, Eve can reverse the permutation among her probes or results and nothing would have been accomplished by the users.

Thirdly, the security proof of MDI-QKD is not spelled out despite its different configuration from usual BB84. In particular, the fact that a third party ``Charles" which performs the required Bell measurement does not have to be trusted by the users does not imply he cannot be fooled by Eve. Detector blinding attacks [7] show that Eve may intercept-resend to take advantage of the detection electronics. Thus, Charles may get results that are skewed by Eve's interception. Whether she may actually be able to influence Charles' detection appears to depend on the exact way and devices he uses to perform the Bell measurement. That no hacking attack has been found does not imply the system has been proved secure against hacking. A proof needs to be provided to show general impossibility. See Section II. The issue of detector representation remains. In addition, MDI-QKD is subject to all the foundational criticisms of [5,6] as well as the physics and system issues raised in this paper, whether it is used with or without decoy states.

\section{CONCLUDING  REMARKS}

Given the history of erroneous claims on QKD security, the seriousness of security claims, and the importance of objective scientific truth, it appears that QKD or any other cryptosystem security proof should be fully spelled out for expert scrutiny. All the underlying assumptions and/or approximations should be clearly identified. Eve's viewpoint needs to be taken seriously in order to include all possible attacks in a general security proof. Full concrete protocol including message authentication should be specified. This situation is especially necessary in quantum cryptography, in which many unfamiliar considerations across different disciplines occur in the subtle context of crypto security. It is easy to make error or overlook issue. Only with detailed scrutiny can a cryptosystem designer or an application user be convinced of the in-principle security of a protocol, which hopefully experimentalists would be able to implement in time. A set of slides is available [35] for a summary of these and related points.

\section*{ACKNOWLEDGMENT}
The work reported in this article was supported by the Air Force Office of Scientific Research and the Defense Advanced Research Project Agency.

\end{document}